\documentclass[letterpaper, 10 pt, conference]{ieeeconf}  

\IEEEoverridecommandlockouts                              

\usepackage{makecell}
\usepackage{booktabs}
\usepackage{multirow} 
\usepackage{graphics} 
\usepackage{caption}
\usepackage{graphicx}
\usepackage{float} 
\usepackage{subcaption}
\usepackage[justification=centering]{caption}
\usepackage{amsmath}

\title{\LARGE \bf
Cross-Domain Transfer Learning Method for Thermal Adaptive Behavior Recognition with WiFi*
}

\author{Zhaohe Lv, Guoliang Zhao, Zhanbo Xu, \textit{Member, IEEE,} Jiang Wu, \textit{Member, IEEE,}\\Yadong Zhou, \textit{Member, IEEE,} and Kun Liu, \textit{Member, IEEE} 
\thanks{*This work was supported in part by National Natural Science Foundation
of China (62122062, 62192755, 62192750, 62192752).}
\thanks{Z. Lv, G. Zhao, Z. Xu, J. Wu, Y. Zhou, K. Liu are with the Moe Klinns Lab and Systems Engineering Institute of Xi’an Jiaotong University, Xi’an, 710049, China (e-mail: zbxu@sei.xjtu.edu.cn).}
}

\begin{document}

\maketitle
\thispagestyle{empty}
\pagestyle{empty}

\begin{abstract}
A reliable comfort model is essential to improve occupant satisfaction and reduce building energy consumption. As two types of the most common and intuitive thermal adaptive behaviors, precise recognition of dressing and undressing can effectively support thermal comfort prediction. However, traditional activity recognition suffers from shortcomings in privacy, cost, and performance. To address the above issues, this study proposes a cross-domain transfer learning method for human dressing and undressing adaptive behavior recognition with WiFi. First, we determine the activity interval by calculating the sliding variance for denoised WiFi signals. Subsequently, short-time Fourier transform and discrete wavelet transform are performed to extract action information on the basis of time-frequency analysis. Ultimately, an efficient 1D CNN pre-trained model is integrated with the SVM algorithm as a hybrid model to enhance the identification robustness in new scenarios. Experiment results show that the hybrid model based on transfer learning provides a more accurate prediction for the adaptative behavior of target subjects, achieving 96.9\% and 94.9\% accuracy in two cases, respectively.

\end{abstract}

\begin{figure*}[htbp]
    \centering
    \includegraphics[width=0.80\linewidth]{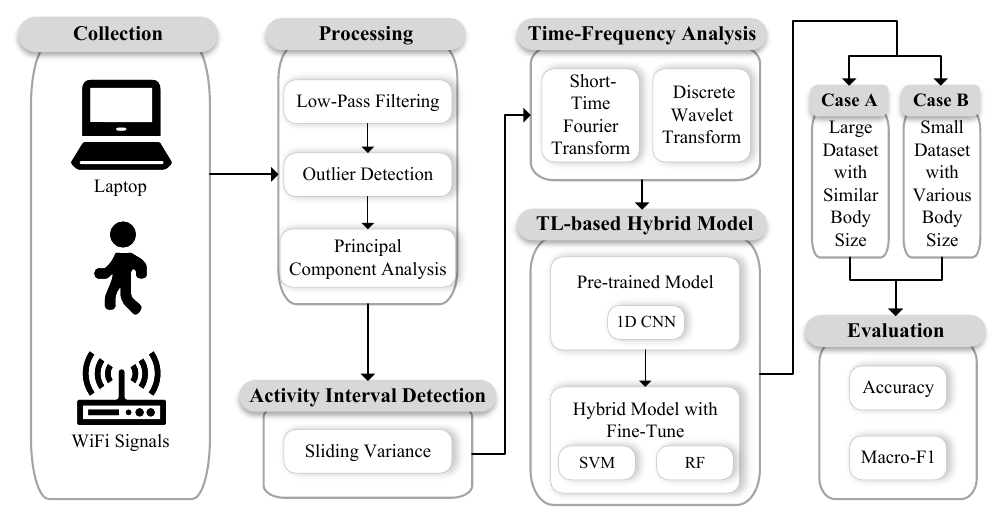}
    \caption{Research flow.}
    \label{progress}
\end{figure*}

\section{INTRODUCTION}

With the continuous development of IoT, human sensing has become the core technology for building human-centered IoT application systems. In particular, human activity recognition is widely used in health care \cite{li2017r}, intelligent security \cite{chung2015applying}, smart homes \cite{wan2015dynamic}, etc. Furthermore, relevant studies have indicated that people spend 90\% of their time indoors \cite{klepeis2001national}. This highlights the need to create a healthy and comfortable environment for occupants where thermal comfort is an essential indicator. When experiencing thermal discomfort, indoor occupants usually react in a way that allows returning to a relaxed state, referred to as thermal adaptive behavior \cite{nicol1998understanding}. Thermal adaptive behavior is a significant expression of thermal preferences \cite{mustapa2016thermal}. Prediction of human thermal adaptive behavior contributes to the development of an accurate thermal comfort model which can provide the occupant with a highly customizable comfort climate zone. One study \cite{ghahramani2014knowledge} shows the integration of personal comfort models into thermostat control to determine the optimal temperature set point for a selected area. The results reveal a 12\% reduction in mean airflow rate in the tested areas while maintaining or improving comfort.

Our work is dedicated to the identification of thermal adaptive behavior, which is roughly related to the following three aspects:

\textbf{Thermal adaptive behavior.} As described in \cite{nicol1998understanding}, human adaptive behaviors can be divided into two categories: thermal control behaviors and self-regulatory behaviors. Adjusting the blinds and turning on the thermostat are examples of thermal control behaviors. Wiping, dressing and undressing are typical self-regulatory behaviors. Studies \cite{gunay2014adaptive} verify the relationship between window control and occupant comfort. Kim et al. \cite{kim2018personal} obtain a median accuracy of 0.73 for comfort prediction by recording the frequency and intensity of using private seats with heating and cooling systems. Liu et al. \cite{liu2021thermal} develop the thermal comfort model based on observation of human self-regulatory behaviors from a field lab, achieving better performance than environmental and physiological factors. While these studies demonstrate the broad utility of thermal adaptive behavior in model development as feedback that does not disturb the occupant, most of them cannot automatically detect actions, which is time-consuming and impractical to apply. 

\textbf{Activity recognition.} Appropriate human sensing technology is essential to construct a personal comfort system. Existing methods for human motion recognition fall into two broad types: vision-based systems \cite{mo2016human} and wearable sensor-based systems \cite{chen2021deep}. Vision-based systems rely on computer vision techniques and cameras to recognize activity; yet, they are sensitive to light and fail to work when obscured, along with privacy issues, rendering them suitable for limited application scenarios. Wearable sensor-based systems facilitate data collection but need to be carried around. In comparison with other technologies, emerging WiFi sensing technology has attracted many researchers due to the benefits of a wide sensing range, non-intrusiveness, and privacy protection. With the help of channel state information (CSI), pioneering attempts such as E-eyes \cite{wang2014eyes} and CARM \cite{wang2015understanding} have been proposed. Nevertheless, wireless signals and their characteristics are highly dependent on the environment in which the subject is located, so that classifiers trained with initial signals in one scenario generally suffer poor performance in another environment.

\textbf{Transfer learning.} Facing the dilemma of insufficient labeled data and poor generalization ability in cross-domain activity recognition, we would like to build a universal prediction model to achieve the expected performance without collecting extensive data for retraining. Transfer learning (TL) better meets this need by solving problems in related domains according to known knowledge. Cook et al. \cite{cook2013transfer} discuss the feasibility of knowledge transfer for sensor-based behavior recognition. With the hidden Markov model, Pan et al. \cite{pan2008transfer} conduct a study on indoor WiFi localization in different spaces. Long et al. \cite{long2017deep} improve the structure of deep networks to reduce differences between various domains through probability distribution adaptations. However, few studies have focused on the transfer strategy integrated with machine learning classifiers.

To overcome such challenges and expressly understand the thermal preferences of individual occupants, We propose a cross-domain transfer learning method for human dressing and undressing adaptive behavior recognition with WiFi, named TL-WiDUR. TL-WiDUR uses CSI provided by commodity WiFi devices for occupant activity recognition, which supports the development of individual thermal comfort models. In particular, dressing and undressing, as the most common thermal adaptive behaviors, are selected to verify the feasibility of TL-WiDUR. In this study, we make the following contributions:
\begin{itemize}
\item We firstly propose a cross-domain transfer learning method to recognize thermal adaptive behavior with WiFi, especially for typical dressing and undressing, which is a significant attempt to apply wireless signals in thermal comfort modeling. On the basis of traditional personal comfort systems, we provide precise thermal preference prediction in terms of improving occupant comfort and building thermal efficiency.
\item To better distinguish between dressing and undressing and other activities, we fully utilize all 30 subcarriers and extracted features from both wavelets and spectrums to reduce the effect of extraneous environmental information carried by the signal. As for low recognition rates in new environments, we apply TL and hybrid models to map the weight vector trained from the primary dataset to the target domain, providing a low-cost solution to alleviate the lack of data quantity and enhance robustness.
\item Using commercial devices to capture CSI, we propose the first WiFi dataset designed explicitly for human thermal adaptive behavior detection with the example of dressing and undressing, offering a critical reference for following research in this field. Within two different rooms, over 4800 activity samples from three subjects are collected to assess the robustness of our model.

\end{itemize}

\section{OVERVIEW OF TL-WiDUR}
The overall process of the study is depicted in Fig. \ref{progress}, and TL-WiDUR consists of two hardware elements: a commercial WiFi access point (AP) and a laptop with a WiFi card. Since environmental changes can affect the wireless signal, noise should be filtered and outliers should be removed to improve the quality of collected data. Besides, the 30 subcarriers of each CSI stream are highly associated with each other, resulting in feature redundancy. We apply principal component analysis (PCA) to detect time-varying correlations among them and extract the first principal component with the highest variance. To enhance recognition accuracy, the activity interval is determined by comparing the sliding variance with a threshold value. In the feature extraction stage, short-time Fourier transform (STFT) and discrete wavelet transform (DWT) are used to capture body movements. Afterward, we explore the possibility of the one-dimensional convolutional neural network (1D CNN) in combination with support vector machine (SVM) and random forest (RF) classifiers to create a hybrid model based on TL. Analysis experiments on the performance in two scenarios are performed to examine the stability and viability of the proposed scheme for adaptive behavior recognition. 

\section{METHODOLOGY}
\subsection{Activity Interval Detection}

\begin{figure}[htb]
	\centering
	\begin{subfigure}{0.49\linewidth}
           \centering
		\includegraphics[width=1\linewidth]{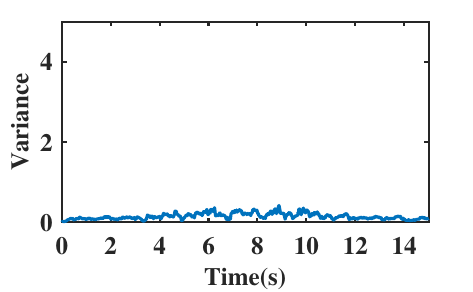}
		\caption{Magnitude variance of static environment}
		\label{var_empty}
	\end{subfigure}
	\begin{subfigure}{0.49\linewidth}
           \centering
		\includegraphics[width=1\linewidth]{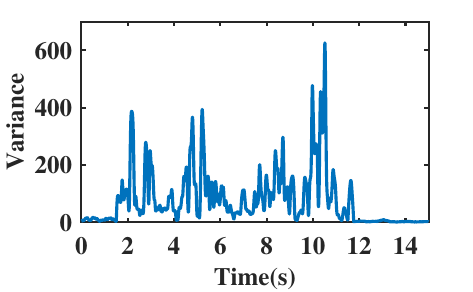}
		\caption{Magnitude variance of dynamic environment}
		\label{var_action}
	\end{subfigure}
	\begin{subfigure}{0.8\linewidth}
           \centering
		\includegraphics[width=1\linewidth]{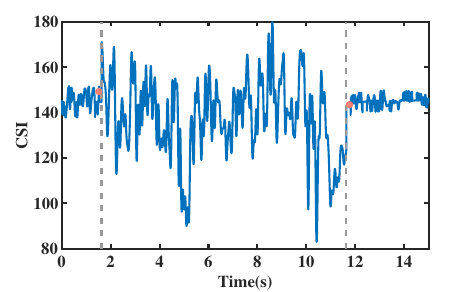}
		\caption{Activity interval detection}
		\label{begin_end}
	\end{subfigure}
	\caption{Magnitude variance and interval detection of different environments.}
	\label{movvar}
\end{figure}

Detecting human activity intervals plays an essential role in enhancing recognition precision. As the person moves around the room, additional paths are introduced by the reflection of the human body, and CSI values will reflect the variations of these paths. Accordingly, the degree of amplitude variation in the CSI stream, defined as sliding variance, can be used to search the beginning and end of the activity. Assuming $x_{j}$ is the $j$th sample in the CSI stream, then variance $\sigma _{j}^{2}$ can be expressed as:
\begin{equation}
\sigma_{j}^{2}=\frac{1}{m} \sum_{j=1}^{m}\left|x_{j}-\mu\right|^{2},\quad (j=1,2, \ldots, n-m-1)
\end{equation}
where $m$ is the interval length of a sliding window, $n$ is the number of packets, and $\mu$ represents the mean value of samples. Fig. \ref{var_empty} and \ref{var_action} illustrate the CSI magnitude variance of the environment at static and with human activity. Intuitively, the curve exhibits significant fluctuations with human presence in the environment. Once the action is completed, sliding variance decreases and smooths out, similar to the static scene. As a result, both two endpoints of each activity are roughly identified from sharp points of the curve. We construct a sliding window for every ten packets in the CSI stream and calculate the variance within it for comparison with the threshold in a static environment. As shown in Fig. \ref{begin_end}, the gray dashed lines denote an actual activity interval, while the red dots represent predicted endpoints. It can be seen that a slight error still exists at the beginning and end of the activity owing to the sensitivity of signals. 

\subsection{Time-Frequency Analysis}
Time-frequency analysis is a major group of methods to deal with non-stationary signals, which can express relevant characteristics of various actions, such as movement speed and time duration. In order to distinguish similar behavior, we combine wavelet layer features along with spectral features as input parameters to the prediction model during feature engineering.
\subsubsection{Short-Time Fourier Transform}
\begin{figure}[htb]
	\centering
	\begin{subfigure}{0.49\linewidth}
           \centering
		\includegraphics[width=1\linewidth]{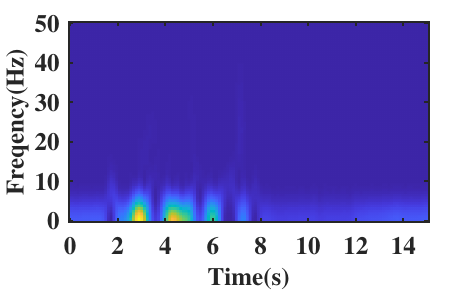}
		\caption{Dressing}
		\label{spec_dress}
	\end{subfigure}
	\begin{subfigure}{0.49\linewidth}
           \centering
		\includegraphics[width=1\linewidth]{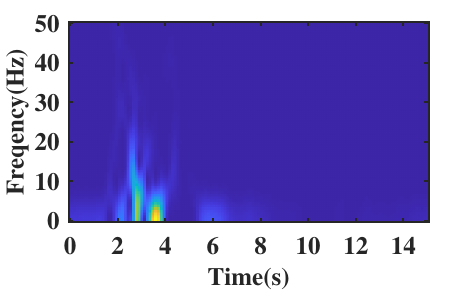}
		\caption{Undressing}
		\label{spec_undress}
	\end{subfigure}
	\caption{Spectrograms for dressing and undressing.}
	\label{spec}
\end{figure}

Based on the Fourier transform, STFT calculates the signal spectrum for each sub-time period separately with a sliding window function over the time-domain signal. $SPEC\left(t,w\right)$ is given by the following equation:
\begin{equation}
SPEC\left(t,w\right) = \left|\int _{ - \infty }^{ + \infty }x\left(\tau \right)h\left(\tau  - t\right)e^{ - jw\tau }d\tau \right|^{2}
\end{equation}
where $x\left(t\right)$ denotes the time domain signal, $h\left(t\right)$ represents the window function, and $SPEC\left(t,w\right)$ is the spectrogram defined by sample $t$ and frequency $w$. Fig. \ref{spec} shows the spectrograms for dressing and undressing. STFT allows capturing of energy distribution which is specific to various activities. For example, the energy band concentrated within 8 Hz as a frequency component of dressing is caused by arm extension and swing while unfolding the clothes. With the help of capturing the movement of body parts, it can be determined whether the activity involves the entire body or only the arms. In addition, the duration of different behaviors varies, with the energy band for dressing ranging from 2 to 8 seconds, while the energy band for undressing mainly lies from 2 to 4 seconds.

\subsubsection{Discrete Wavelet Transform}
\begin{figure}[htb]
	\centering
	\begin{subfigure}{0.49\linewidth}
           \centering
		\includegraphics[width=1\linewidth]{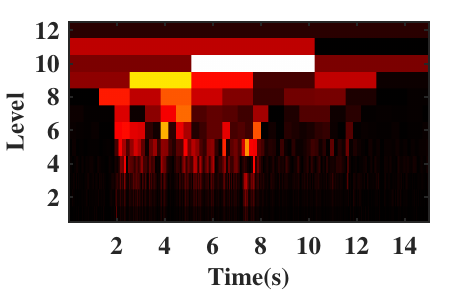}
		\caption{Dressing}
		\label{wave_dress}
	\end{subfigure}
	\begin{subfigure}{0.49\linewidth}
           \centering
		\includegraphics[width=1\linewidth]{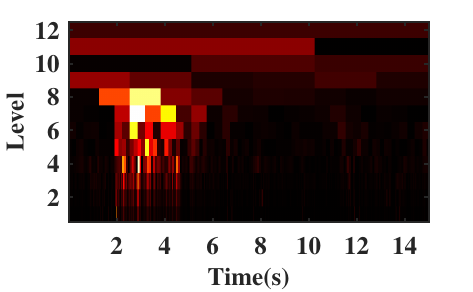}
		\caption{Undressing}
		\label{wave_undress}
	\end{subfigure}
	\caption{DWT of time-series for dressing and undressing.}
	\label{wave}
\end{figure}

DWT enables us to extract frequencies with various resolutions on multiple time scales, providing higher frequency resolution in low-frequency parts of the signal. In this case, different levels of energy are calculated, each corresponding to a range of frequencies. For instance, level 1 expresses the frequency range of 25$\sim$50 Hz, corresponding to the movement speed of 1.5625$\sim$3.125 m/s in the 2.4 GHz band, then level 2 decays exponentially to half the frequency range of level 1, i.e., 12.5$\sim$25 Hz, which corresponds to 0.78125$\sim$1.5625 m/s. Fig. \ref{wave} shows the wavelet compression diagrams for dressing and undressing, where high brightness indicates increased energy density. We can find that the high energy region of dressing moves from level 3 to level 11 in 2 to 10 seconds, while that of undressing moves from level 2 to level 8 in 2 to 5 seconds. Thus, the characteristics of dressing and undressing are reflected at different decomposition levels, which is easy to distinguish them from other behaviors.

\subsection{TL-based Hybrid Model}
For activity recognition in new environments, we adopt TL for knowledge migration to avoid expensive data annotation and explore the cross-domain generalization capability. In the case of uneven sample distribution, the pre-trained model developed should be appropriate for discovering general features from the provided dataset and supplying precise prediction results. Prior studies have indicated that convolutional neural networks maintain a lightweight model with strong nonlinear capabilities and robustness, reducing the difficulty of optimizing structural parameters. Therefore, we compress and splice the collected 2D features into a 1D vector input and apply the widely used feature extraction algorithm 1D CNN as a pre-trained model. The predictor is designed as a 6-layer structure, where three convolutional layers are used to detect potential local information, and the remaining fully connected layers integrate features to complete classification.

\begin{figure}[htb]
    \centering
    \includegraphics[width=1\linewidth]{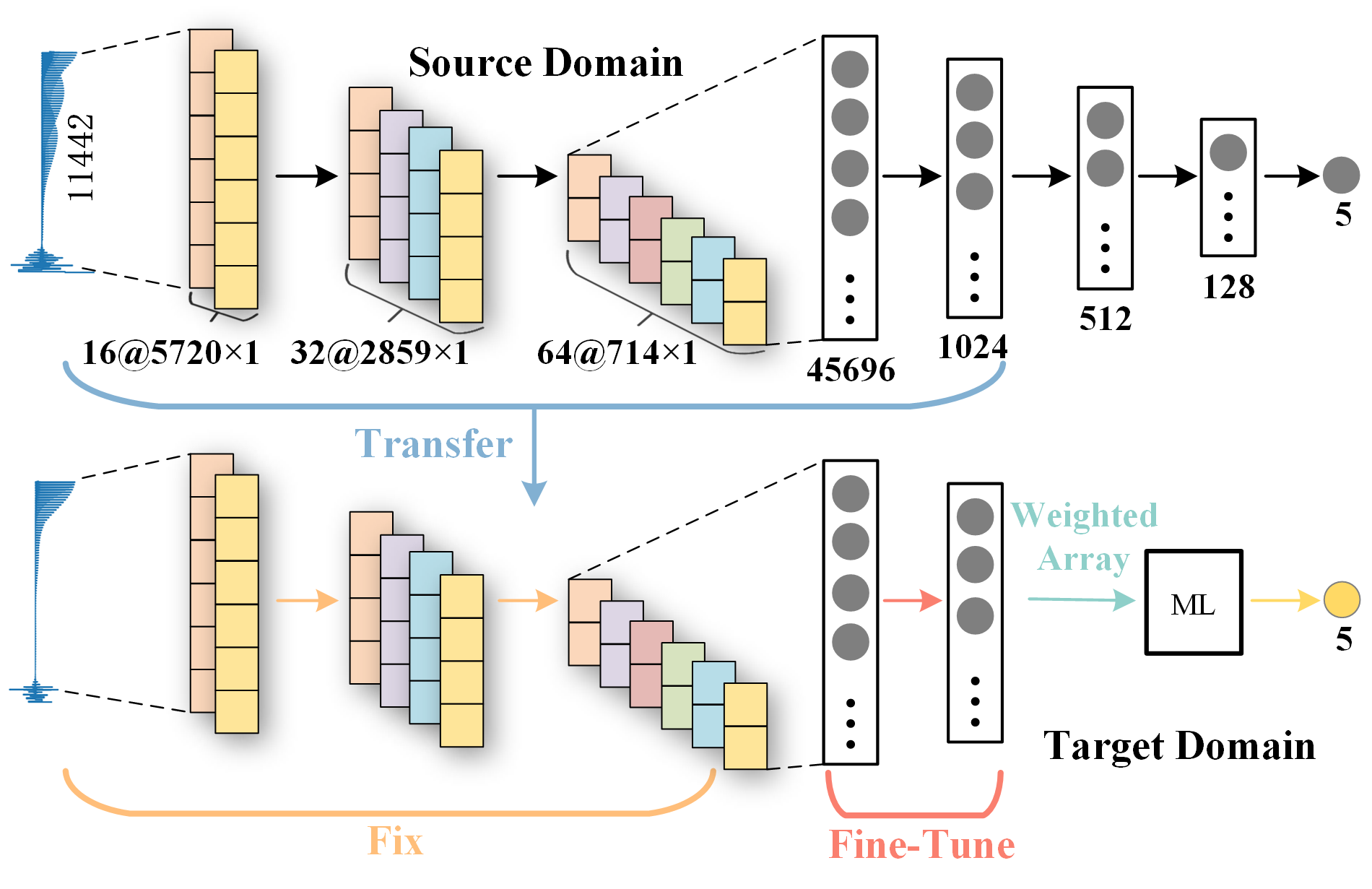}
    \caption{1D CNN hybrid model based on transfer learning.}
    \label{net}
\end{figure}

In practice, target domains usually suffer from little data quantity, unbalanced category labels, and a lack of resources. In this study, a hybrid algorithm fusing deep learning (DL) with traditional machine learning (ML) is utilized to fulfill these considerations. Due to the improvement in prediction results, hybrid models have gained the interest of many researchers. DL algorithms have a solid capability to extract features automatically from structured or unstructured datasets, while ML algorithms guarantee better results in classification with lower underfitting/overfitting. Accordingly, we propose a hybrid model based on TL to ensure maximum prediction performance while transferring knowledge to the target subject, as shown in Fig. \ref{net}. In general, lower layers of the neural network usually contain more generic features as a result of limited perceptual fields. Of all 1D CNN layers, the convolutional layers are frozen, while the first hidden layer of remaining dense layers is intercepted and fine-tuned to reduce feature dimensions. Eventually, the extracted weighted array is reused as input for SVM and RF to classify individual activities as prediction probabilities.

\section{EXPERIMENTATION AND EVALUATION}
The implementation and detailed performance of TL-WiDUR are described in this section.
\subsection{Experiment Methodology}
During the experiment, we implement TL-WiDUR on a Thinkpad X200 laptop equipped with a 5300 WiFi NIC and test it using TP-Link AC1200 as a commercial 801.11ac AP. The laptop is running Ubuntu 14.04 LTS with 4.2.0 kernel. All experiments in this study are performed in the 2.4 GHz band. The packet transmission rate is set to 200 pkts/s. For each packet, we record CSI measurements from 30 subcarriers, which describe fine-grained physical layer properties of the communication channel.

\begin{figure}[htb]
	\centering
	\begin{subfigure}{0.7\linewidth}
           \centering
		\includegraphics[width=1\linewidth]{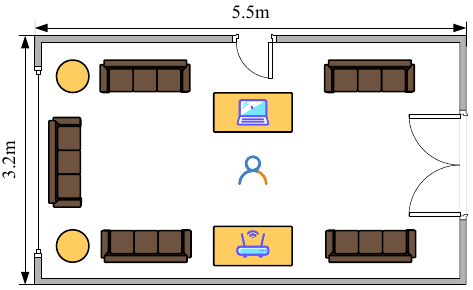}
		\caption{Meeting room}
		\label{meeting}
	\end{subfigure}
	\begin{subfigure}{0.7\linewidth}
           \centering
		\includegraphics[width=1\linewidth]{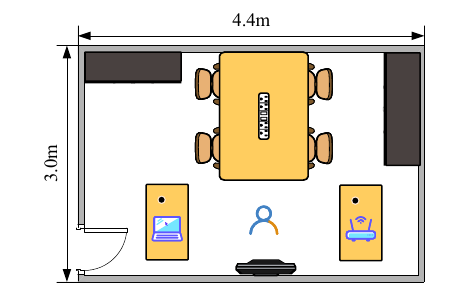}
		\caption{Office}
		\label{office}
	\end{subfigure}
	\caption{Layouts of two experimental environments.}
	\label{room}
\end{figure}

To fully analyze the feasibility of TL-WiDUR, we collect action datasets from three subjects (two males and one female) aged approximately 24. Subjects A and B are adult males of similar height and body shape, while subject C is a shorter female. Fig. \ref{room} depicts the two experimental environments where all three subjects are located. The indoor scenario where subject A located is a typical meeting room with length and width of 5.5m and 3.2m, respectively. In contrast, the environment for collecting datasets from both subjects B and C is a small office with a length and width of 4.4m and 3.0m. All devices are deployed at the height of 70cm, where subjects of varying heights can perform the actions easily. Table \ref{dataset} shows the sample size for different activities of each subject. The maximum sample size is collected from subject A, while the smallest is obtained from subject C in the target domain.

\begin{table}[htb] 
\caption{Distribution of data collected from each subject}
\label{dataset}
\centering
\setlength{\tabcolsep}{7pt}
\begin{tabular*}{\linewidth}{ccccccc} 
\toprule 
\multicolumn{1}{c}{}& \multicolumn{6}{c}{Labels}\\
\cmidrule(lr){2-7}
\multicolumn{1}{c}{}&Empty&Sit&Dress&Undress&Other&Total\\  
\midrule
\multicolumn{1}{c}{Subject A}&848&852&540&541&854&3641\\   
\multicolumn{1}{c}{Subject B}&150&150&96&96&150&642\\
\multicolumn{1}{c}{Subject C}&150&134&79&75&150&588\\
\bottomrule 
\end{tabular*}
\end{table}

\subsection{Overall Performance}
\subsubsection{Performance Comparison of Pre-trained Models}
In order to validate the classification performance of the pre-trained model, we compare the accuracy and F1 score of 1D CNN with Deep Neural Network (DNN), long short term memory (LSTM), extreme gradient boosting (XGBoost), and logistic regression (LR). The grid search method is applied to tune the hyperparameters of each algorithm. We use the entire dataset of subject A, with 90\% of the samples divided into a training set and the rest for testing. Macro-F1 is considered one of the evaluation metrics and can be calculated by the following equation: 
\begin{equation}
Macro-F1=\frac{1}{C}\sum F1_{i} 
\end{equation}
where $C$ is the number of classes and $F1_{i}$ is the F1 score of the $i_{th}$ class. Ten experiments are conducted to verify the stability and validity of each prediction. As depicted in Table \ref{pre_trained}, 1D CNN obtain the best performance among DL or ML-based prediction models with an accuracy of 96.7\% and an F1 score of 97.3\%. DNN and LSTM rank inferior to the former, while LR and XGBoost, two representative ML algorithms, are inefficient in handling complex data and fail to achieve satisfying results. Therefore, we choose 1D CNN as the pre-trained extractor to transfer knowledge about adaptive behavior in this study.

\begin{table}[htb] 
\caption{Comparison of prediction methods}
\label{pre_trained}
\centering
\begin{tabular}{cccc} 
\toprule 
\multicolumn{1}{c}{\multirow{2}{*}{Model}}& &\multicolumn{2}{c}{Metric}\\
\cmidrule(lr){3-4}
\multicolumn{1}{c}{}&&Accuracy&Macro-F1\\  
\midrule
\multicolumn{1}{c}{\textbf{1D CNN}}&&\textbf{0.967}&\textbf{0.973}\\   
\multicolumn{1}{c}{DNN}&&0.962&0.965\\
\multicolumn{1}{c}{LSTM}&&0.956&0.962\\
\multicolumn{1}{c}{XGBoost}&&0.929&0.939\\
\multicolumn{1}{c}{LR}&&0.907&0.921\\
\bottomrule 
\end{tabular}
\end{table}

\subsubsection{Evaluation of TL-based Adaptive Behavior Recognition}
Two cases are selected to test the applicability of TL-WiDUR. In the first case, the dressing and undressing adaptive behavior of subject B is predicted in different indoor environments with similar body sizes and relatively large datasets. What’s more, the TL-based model is further evaluated in the second case with clearly different body sizes and no large datasets. These two cases are considered to reflect realistic conditions, corresponding to the best and worst application scenarios. With the robust perception of dressing and undressing, we can capture subjects' satisfaction with the current environment to build various thermal comfort systems. Table \ref{TL} summarizes the prediction results of four different models for target subjects B and C. It is confirmed that the CNN-SVM hybrid model based on TL (TL-based CNN-SVM) exhibits the best accuracy and F1 score, followed by the CNN-RF hybrid model based on TL (TL-based CNN-RF) and 1D CNN model based on TL (TL-based CNN), both of which perform similarly. In the case of subject B, the TL-based CNN-SVM produces the predictive performance of 96.9\% in accuracy and 97.1\% in F1 score, being improved by 10.7\% and 10.6\% over the CNN standalone model and 1.5\% and 1.4\% over the TL-based CNN in both evaluation metrics.

\begin{table}[htb] 
\caption{Predictive performance of four different models for adaptive behavior of target subjects}
\label{TL}
\centering
\setlength{\tabcolsep}{6.5pt}
\begin{tabular*}{\linewidth}{cccccc} 
\toprule 
\multicolumn{1}{c}{\multirow{2}{*}{Case}}& \multicolumn{1}{c}{\multirow{2}{*}{Metric}}&\multicolumn{4}{c}{Model}\\
\cmidrule(lr){3-6}
\multicolumn{2}{c}{}&CNN&\makecell{TL-based\\CNN}&\makecell{\textbf{TL-based}\\\textbf{CNN-SVM}}&\makecell{TL-based\\CNN-RF}\\  
\midrule
\multicolumn{1}{c}{\multirow{2}{*}{A→B}}&Accuracy&0.862&0.954&\textbf{0.969}&0.954\\   
\multicolumn{1}{c}{}&	Macro-F1&0.865&0.957&\textbf{0.971}&0.955\\
\multicolumn{1}{c}{\multirow{2}{*}{A→C}}&Accuracy&0.831&0.915&\textbf{0.949}&0.915\\
\multicolumn{1}{c}{}&	Macro-F1&0.809&0.911&\textbf{0.948}&0.911\\
\bottomrule 
\end{tabular*}
\end{table}

\begin{table*}[htb] 
\caption{Evaluation metrics across different input parameters}
\label{parameter}
\centering
\begin{tabular*}{0.95\linewidth}{cccccccccccc} 
\toprule 
\multicolumn{1}{c}{\multirow{3}{*}{\makecell{Source\\ subject}}}&\multicolumn{1}{c}{\multirow{3}{*}{\makecell{Target\\ subject}}}&\multicolumn{1}{c}{\multirow{3}{*}{Parameter}}& &\multicolumn{8}{c}{Model}\\
\cmidrule(lr){5-12}
\multicolumn{4}{c}{}& \multicolumn{2}{c}{DNN}&\multicolumn{2}{c}{CNN}&\multicolumn{2}{c}{TL-based CNN}&\multicolumn{2}{c}{TL-based CNN-SVM}\\  
\cmidrule(lr){5-6}\cmidrule(lr){7-8}\cmidrule(lr){9-10}\cmidrule(lr){11-12}
\multicolumn{4}{c}{}&Accuracy&Macro-F1&Accuracy&Macro-F1&Accuracy&Macro-F1&Accuracy&Macro-F1\\  
\midrule
\multicolumn{1}{c}{\multirow{9}{*}{A}}&\multicolumn{1}{c}{\multirow{3}{*}{A}}&STFT&&0.918&0.928&0.937&0.946&/&/&/&/\\   
\multicolumn{2}{c}{}&	DWT&&0.841&0.854&0.896&0.911&/&/&/&/\\
\multicolumn{2}{c}{}&	\textbf{STFT DWT}&&\textbf{0.962}&\textbf{0.965}&\textbf{0.967}&\textbf{0.973}&/&/&/&/\\
\multicolumn{1}{c}{}&\multicolumn{1}{c}{\multirow{3}{*}{B}}&STFT&&0.677&0.716&0.646&0.688&0.923&0.919&0.954&0.955\\   
\multicolumn{2}{c}{}&	DWT&&0.538&0.455&0.831&0.852&0.938&0.935&0.954&0.951\\
\multicolumn{2}{c}{}&	\textbf{STFT DWT}&&\textbf{0.785}&\textbf{0.792}&\textbf{0.862}&\textbf{0.865}&\textbf{0.954}&\textbf{0.957}&\textbf{0.969}&\textbf{0.971}\\
\multicolumn{1}{c}{}&\multicolumn{1}{c}{\multirow{3}{*}{C}}&STFT&&0.610&0.464&0.678&0.617&0.881&0.883&0.932&0.929\\   
\multicolumn{2}{c}{}&	DWT&&0.576&0.498&0.797&0.791&0.881&0.874&0.915&0.904\\
\multicolumn{2}{c}{}&	\textbf{STFT DWT}&&\textbf{0.644}&\textbf{0.537}&\textbf{0.831}&\textbf{0.809}&\textbf{0.915}&\textbf{0.911}&\textbf{0.949}&\textbf{0.948}\\
\bottomrule 
\end{tabular*}
\end{table*}

Meanwhile, as for subject C, the TL-based CNN-SVM improved accuracy by 11.8\% and F1 score by 13.9\% compared to the CNN standalone model, as well as 3.4\% and 3.7\% higher than the TL-based CNN, which indicates TL could be a convenient way to deal with insufficient training data. In addition, the prediction model with a hybrid strategy can optimize performance in the target domain. Notably, RF has no worthwhile contribution due to its unsuitability for high-dimensional sparse samples. By contrast, the final decision function of SVM is determined by only a small set of support vectors, exhibiting better results for small sample sizes and making the model more robust to a certain degree. To sum up, TL allows for superior predictive performance by fine-tuning behavior characteristics from the source domain, while the SVM classifier strengthens the generalization ability of TL-WiDUR.

\subsection{Ablation Study}
Moreover, we further analyze the impact of various time-frequency features selected on model performance. Table \ref{parameter} summarizes the predictive results with different input parameters in various domains. Without pre-training, it is apparent that the maximum prediction accuracy can be obtained by applying features extracted from both STFT and DWT as inputs, especially for small sample datasets. Compared to one parameter only, the improvement ratio in evaluation metrics of 1D CNN is over 2\% for all subjects. What's more, the DNN improves accuracy by 24.7\% and F1 score by 33.7\%, respectively, compared to the worst prediction case. A basically consistent trend is produced in transferring knowledge to target subjects B and C, where both parameters bring a significant improvement on model performance. Overall, as two types of complementary and standard time-frequency analysis methods, STFT and DWT can maximize the efficiency of knowledge transfer and improve the reusability of recognition models owing to solid robustness.

\section{CONCLUSION AND FUTURE WORK}
Predicting dressing and undressing adaptive behavior helps to understand individual thermal preferences and improve occupancy satisfaction in a rapid, convenient, effective, and accurate way. However, recognition in unfamiliar environments without additional hardware devices is challenging. Due to the limited quality of samples and the scale of datasets, existing perception methods have significant restrictions in applications. In this study, making full use of the time-frequency features, we firstly propose TL-WiDUR. The method is designed to transfer knowledge from datasets collected in different indoor scenarios, even if data of target subjects are inadequate and labels are unevenly distributed. We evaluate the stability of the prediction model in two different ways, which can reach average accuracies as high as over 94\% and 96\% in target domain datasets of subjects B and C with a single AP and only one WiFi device, respectively. In addition, we conduct supplementary experiments to explore the effects generated by the time-frequency parameters setting. Our results demonstrate that the proposed scheme remarkably improves perceptual performance in new scenarios and has positive significance in constructing thermal comfort models.

In practical circumstances, TL-WiDUR is currently limited to single-person scenarios as signal propagation suffers from interference in complex environments leading to attenuation and ambiguity of the collected features. Further realizing high-precision prediction in multi-person scenarios is the center of future work.










\bibliographystyle{IEEEtran}
\bibliography{ieeeexample}

\end{document}